\documentclass[12pt]{article}
\usepackage{jhep-mod}
\usepackage{bm}
\usepackage{amssymb, amsmath, amsthm, physics}
\usepackage{mathrsfs}
\usepackage{hyperref}
\usepackage{multicol}
\usepackage{float, array, tabu}
\usepackage[tableposition=top]{caption}
\usepackage{tocloft}% http://ctan.org/pkg/tocloft
\usepackage[utf8]{inputenc}
\usepackage{enumerate}
\usepackage{bigints}
\usepackage{appendix}
\usepackage{graphicx}
\usepackage{float}
\usepackage{tikz}
\usepackage{setspace}
\usepackage{cancel}
\definecolor{purple}{rgb}{1,0,1}
\definecolor{lime}{HTML}{A6CE39} % needs xcolor
\newcommand{\blue}[1]{{\color{blue} #1}}

\definecolor{lime}{HTML}{A6CE39}
\newcommand{\orcidicon}{%
	\begin{tikzpicture}
	\draw[lime, fill=lime] (0,0) 
		circle [radius=0.16] 
		node[white] {{\fontfamily{qag}\selectfont \tiny ID}};
	\draw[white, fill=white] (-0.0625,0.095) 
		circle [radius=0.007];
	\end{tikzpicture}
	\hspace{-5mm}
}
\newcommand\orcidMatt{{\href{https://orcid.org/0000-0003-1088-6485}{\orcidicon}}}
\begin{document}
%========================================================
\def\O{{\mathcal{O}}}
%========================================================
\title{%\vspace{-75pt}
\huge{
%\leftline{\red{*** DRAFT *** DRAFT *** DRAFT ***}}
\leftline{Reconsidering  maximum luminosity}
%\leftline{\red{*** DRAFT *** DRAFT *** DRAFT ***}}
}
}
%========================================================
%========================================================
\author{
\Large
Aden Jowsey {\sf{and}} Matt Visser\orcidMatt$^*$}
%========================================================
%========================================================
%========================================================
%========================================================
\affiliation{School of Mathematics and Statistics, Victoria University of Wellington, \\
\null\qquad PO Box 600, Wellington 6140, New Zealand.}
%========================================================
%========================================================
\emailAdd{aden.jowsey@sms.vuw.ac.nz}
\emailAdd{matt.visser@sms.vuw.ac.nz}
%========================================================
%========================================================
\parindent0pt
\parskip7pt

\abstract{

\noindent
The suggestion that there is a maximum luminosity (maximum power) in nature has a long and somewhat convoluted history. Though this idea is commonly attributed to Freeman Dyson, he was  actually much more circumspect in his views.
What is certainly true is that dimensional analysis shows that the speed of light and Newton's constant of gravitation can be combined to define a quantity $P_* = {c^5\over G_N}$ with the dimensions of luminosity (equivalently, power). 
Then in \emph{any} physical situation we \emph{must} have $P_\mathrm{physical} = \wp \; P_*$, where the quantity  $\wp$ is some dimensionless function of dimensionless parameters. 
This has lead some authors to suggest a maximum luminosity/maximum power conjecture. 
Working within the framework of standard general relativity, we will re-assess this conjecture, paying particular attention to the extent to which various examples and  counter-examples are physically reasonable.  
We focus specifically on Vaidya spacetimes, and on an evaporating version of Schwarzschild's constant density star. For both of these spacetimes luminosity can be arbitrarily large. 
We argue that any luminosity bound must depend on delicate internal features of the radiating object.

\bigskip
\noindent
{\sc Date:} Tuesday 30 March 2021; \LaTeX-ed \today


\bigskip
$^*$ Corresponding author.

\bigskip
\blue{
Honourable mention in the 2021 Gravity Research Foundation essay contest.
}
}

%========================================================
{\enlargethispage{50pt}
\vspace{-30pt}
\setlength{\cftbeforesecskip}{3pt}
\maketitle}
%========================================================
\def\tr{{\mathrm{tr}}}
\def\diag{{\mathrm{diag}}}
%=======================================
\def\H{{\scriptscriptstyle{\mathrm{H}}}}
\def\O{{\mathcal{O}}}
%-----------------------------------------------------------------------------
\def\lint{\hbox{\Large $\displaystyle\int$}} %needs \usepackage{amssymb}
\def\hint{\hbox{\Huge $\displaystyle\int$}}  %needs \usepackage{amssymb}
%-----------------------------------------------------------------------------
%=======================================
%\clearpage
%=======================================
\section{Introduction}\label{S:intro}
%=======================================

One starts by noting that in (3+1) dimensions the quantity 
\begin{eqnarray}
    \label{eq:P_*} P_* =\frac{c^5}{G_N} = 1 \hbox{ Dyson} \approx 3.6\times10^{52} \hbox{ W}
\end{eqnarray}
has the dimensions of luminosity (equivalently, power). Here $c$ is the speed of light in vacuum, and $G_N$ is Newton's gravitational constant. Thereby, using straightforward dimensional analysis,  \emph{any} physical luminosity can \emph{always} be written in the form 
\begin{equation}
P_\mathrm{physical} = \wp\; P_*,
\end{equation}
where the quantity $\wp$ is some dimensionless function of dimensionless parameters. 

The suggestion that $\wp$ is bounded, that is, $\wp=\O(1)$, is commonly misattributed to Freeman Dyson. See reference~\cite{Dyson}. Some relevant historical background is  reported in reference~\cite{Barrow:2017}; see particularly footnote 5  in reference~\cite{Barrow:2017}:
\begin{quotation}
``It is not true that I proposed the formula $c^5/G$ as a luminosity limit for anything. I make no such claim. Perhaps this notion arose from a paper that I wrote in 1962 with the title, ``Gravitational Machines'', published as Chapter 12 in the book, ``Interstellar Communication'' edited by Alastair Cameron, [New York, Benjamin, 1963]. Equation (11) in that paper is the well-known formula $128V^{10}/5Gc^5$ for the power in gravitational waves emitted by a binary star with two equal masses moving in a circular orbit with velocity $V$. As $V$ approaches its upper limit $c$, this gravitational power approaches the upper limit $128c^5/5G$. The remarkable thing about this upper limit is that it is independent of the masses of the stars. It may be of some relevance to the theory of gamma-ray bursts.''\footnote{Dyson's article was, in its original essay form, also awarded 4th prize in the 1962 Gravity Research Foundation essay contest, now some 59 years ago.}$^{,}$\footnote{The fact that $128/5 > 25 \gg 1$ should perhaps encourage a certain amount of caution regarding any precise numerical bound being placed on the dimensionless number $\wp$.
}\\
\null\hfill{---Freeman Dyson}
\end{quotation}

Indeed, the fact that $P_* =\frac{c^5}{G_N}$ has units of power can be traced back to the 1880s, to the development of the classical Stoney units, which pre-date Planck units by some 20 years~\cite{Stoney1, Stoney2, Stoney3}. The classical Stoney units use $G_N$, $c$,
and Coulomb's constant $e^2\over4\pi\epsilon_0$, (instead of $G_N$, $c$, and Planck's constant $\hbar$), to set up a universal physically motivated system of units.  

Some of the classical Stoney units are equal to the corresponding better-known quantum-inspired Planck units --- those where the factors of Coulomb's constant or $\hbar$ cancel. 
Specifically we have  $P_*= P_\mathrm{Planck}= P_\mathrm{Stoney}=\frac{c^5}{G_N}$. 
Similarly  we have natural units of force $F_* = F_\mathrm{Planck} =F_\mathrm{Stoney}=\frac{c^4}{G_N}$,  mass-loss-rate $(\dot m)_* = (\dot m)_\mathrm{Planck}= (\dot m)_\mathrm{Stoney}=\frac{c^3}{G_N}$, 
and mass-per-unit-length $(m')_* = (m') _\mathrm{Planck} = (m')_\mathrm{Stoney}= \frac{c^2}{G_N}$. 
Based ultimately on simple dimensional analysis, any one of these natural units might be used to advocate for a maximality conjecture: maximum luminosity~\cite{Dyson, Barrow:2017, Hogan:1999, Schiller:2005, Cardoso:2018, Bruneton:2013}, maximum force~\cite{Gibbons:2002,Schiller:1997,Schiller:2005, Bruneton:2013, Barrow:2014,Barrow:2020,Ong:2018}, maximum mass-loss-rate, or maximum mass-per-unit-length. 
We have recently argued for a certain amount of caution regarding the conjectured bound on maximum force~\cite{Jowsey:2020}, and in this essay we will now turn attention to the maximum luminosity conjecture. 

Now it is certainly true that in very many specific situations~\cite{Gibbons:2002,Schiller:1997,Barrow:2014,Barrow:2020} explicit calculations yield $\wp \leq {1\over4}$, though sometimes numbers such as $\wp\leq {1\over2}$ also arise~\cite{Hogan:1999}. 
Specifically,  consider strong, medium, and weak versions of the maximum luminosity conjecture:
 \begin{enumerate}
 \itemsep-1pt
\item {Strong form:} \quad\quad $\wp\leq {1\over4}$.    
\item {Medium form:} \quad $\wp\leq {1\over2}$.       
 \item {Weak form:} \quad\quad $\wp = \O(1)$.
 \end{enumerate}
 The question we wish to address is whether or not these conjectured bounds are truly universal. 
 See particularly the cautionary comments in~\cite{Cardoso:2018}.
 
%========================================================
\section{Vaidya spacetimes}\label{C:Vaidya}
%========================================================
 
 Let us first consider Vaidya's spacetime, which is most typically interpreted as the \emph{exterior} geometry of a shining star~\cite{Vaidya:1951, Vaidya:1999a, Vaidya:1999b}.
We find it convenient to use $(t,r,\theta,\phi)$ coordinates,  set $c\to 1$, and to represent the line element in Kerr--Schild form:
\begin{equation}
ds^2 = - dt^2 + dr^2 + r^2(d\theta^2+\sin^2\theta\, d\phi^2) + {2 G_N\; m(t-r)\over r  } (dr- dt)^2.
\end{equation}
Equivalently
\begin{equation}
g_{ab} = \eta_{ab} +  {2G_N \; m(t-r)\over r}\;   \ell_a \ell_b,
\end{equation}
where  the vector $\ell_a=(-1,1;0,0)_a$ is null both with respect to the flat metric $\eta_{ab}$ and the physical metric $g_{ab}$.  It is then a straightforward and standard calculation to check that
\begin{equation}
G_{ab} =- {2 G_N \; \dot m(t-r)\over r^2} \; \ell_a \ell_b.
\end{equation}
Applying the Einstein equations, $G_{ab} = 8\pi G_N \; T_{ab}$, one has
\begin{equation}
T^{ab} = - {\dot m(t-r)\over 4 \pi r^2} \, \ell^a \ell^b.
\end{equation}
Checking that 
\begin{equation}
R_{\hat\theta\hat\phi\hat\theta\hat\phi} = {2 G_N m(t-r)\over r^3},
\end{equation}
verifies that the function $m(t-r)$ is indeed the Misner--Sharp quasilocal mass~\cite{Misner-Sharp}.
Then the luminosity, as seen by a static observer at some fixed value $r_*$ of the radial coordinate, is simply
\begin{equation}
P(t; r_*) = \hbox{(flux)}\times\hbox{(area)} = -\left({\dot m(t-r_*)\over 4 \pi r_*^2}\right) \times (4\pi r_*^2) =  -\dot m(t-r_*).
\end{equation}
(Positive luminosity corresponds to mass loss by the central object.)

 Reinstating SI units
 \begin{equation}
P(t; r_*) =  -\dot m(t-r_*/c)\; c^2 .
\end{equation}
In terms of the Schwarzschild radius $r_\mathrm{Schwarzschild}(t,r_*)=2 G_N m(t-r_*/c)/c^2$ one has
 \begin{equation}
P(t; r_*) = -{1\over2}  {c^5\over G_N}\;  {\dot r_\mathrm{Schwarzschild}(t-r_*/c)\over c} = -{1\over 2} \;P_* \; {\dot r_\mathrm{Schwarzschild}(t-r_*/c)\over c} 
\end{equation}
That is
\begin{equation}
\wp(t; r_*) =- {1\over 2} \;  {\dot r_\mathrm{Schwarzschild}(t-r_*/c)\over c}.
\end{equation}
The point is that the mass function $m(t-r)$, and consequently $r_\mathrm{Schwarzschild}(t-r_*/c)$, is completely arbitrary --- the exterior Vaidya spacetime by itself places no constraint on the luminosity.
(That is, there is no plausibly justifiable physics reason for  demanding $|\dot r_\mathrm{Schwarzschild}(t-r_*/c)|< c$.)
Consequently, the only hope one has for possibly deriving a general purpose maximum luminosity bound must depend on the \emph{interior} geometry of the source, 
not the \emph{exterior} geometry of the radiating object. 

%\clearpage
%========================================================
\section{Schwarzschild's constant density star: Evaporating version}\label{C:Schwarzschild}
%========================================================

As a first crude model for the  \emph{interior} geometry of the source, let us consider a time-dependent evaporating version of Schwarzschild's constant density star. 
(For general background, see the Delgaty--Lake review~\cite{Delgaty-Lake}.)
In the usual Schwarzschild curvature coordinates (Hilbert--Droste coordinates) take:
\begin{eqnarray}
ds^2 &=&
- {\left(3  \sqrt{1 - {8\pi\over3} G_N \rho_* \; r_s(t)^2}  - \sqrt{1 - {8\pi\over3} G_N \rho_* \; r^2} \right)^2\over 4}  \; dt^2 
\nonumber\\ && \qquad\qquad
+ {dr^2\over 1 - {8\pi\over3} G_N \rho_* \; r^2} + r^2(d\theta^2+\sin^2\theta\, d\phi^2).
\end{eqnarray}
Here $\rho_*$ will indeed prove to be the (constant) mass density of the source, while $r_s(t)$ will prove to be the time-dependent radius of the source.

%\clearpage
\enlargethispage{20pt}
A brief calculation yields the orthonormal components of the Einstein tensor:
\begin{eqnarray}
G_{\hat t\hat t} &=& 8\pi G_N \rho_*;
\\[7pt]
G_{\hat r\hat r}&=&G_{\hat \theta\hat \theta}=G_{\hat \phi\hat \phi} 
\\
&=&
8\pi G_N \rho_* \; { 4  \sqrt{1 - {8\pi\over3} G_N \rho_* \; r^2}  \sqrt{1 - {8\pi\over3} G_N \rho_* \; r_s(t)^2} 
- 4 + {8\pi G_N\rho_*\over 3} (3 r_s(t)^2 + r^2)
\over
\left(3  \sqrt{1 - {8\pi\over3} G_N \rho_* \; r_s(t)^2}  - \sqrt{1 - {8\pi\over3} G_N \rho_* \; r^2} \right)^2
}\qquad
\nonumber
\end{eqnarray}
Imposing the Einstein equations, we see that $\rho=\rho_*$ is indeed a constant as advertised.

\clearpage
 Furthermore the internal pressure is now explicitly time-dependent
\begin{equation}
p(r,t) = 4 \rho_* \; {  \sqrt{1 - {8\pi\over3} G_N \rho_* \; r^2}  \sqrt{1 - {8\pi\over3} G_N \rho_* \; r_s(t)^2} 
- 1 + {8\pi G_N\rho_*\over 3} \left(3 r_s(t)^2 + r^2\over4\right)
\over
\left(3  \sqrt{1 - {8\pi\over3} G_N \rho_* \; r_s(t)^2}  - \sqrt{1 - {8\pi\over3} G_N \rho_* \; r^2} \right)^2
}.\qquad
\end{equation}
Note that the pressure does indeed go to zero as $r\to r_s(t)$. 
Physically this geometry corresponds to a non-moving core, $r<r_s(t)$, with the outer layers, $r>r_s(t)$ being blown off. Granted this is a crude model for the interior structure of an evaporating star, but it is good enough to get the main issues across.

The Misner--Sharp quasi-local mass in the bulk is simply~\cite{Misner-Sharp}
\begin{equation}
m(r)=  {4\pi\over 3} \rho_* \; r^3. 
\end{equation}
while the total  Misner--Sharp quasi-local mass
evaluated at the surface $r_s(t)$ is simply
\begin{equation}
m(t) =  {4\pi\over 3} \rho_* \; r_s(t)^3. 
\end{equation}
(This Misner--Sharp quasi-local mass
evaluated at the surface has to match the Misner--Sharp quasi-local mass for the exterior Vaidya spacetime.)

Thence the luminosity is
\begin{equation}
P(t) = -\dot m(t) =  -3 m(t) \; {\dot r_s(t)\over r_s(t)}.
\end{equation}
Reinstating SI units
\begin{eqnarray}
P(t) &=& -\dot m(t) c^2=  -3 m(t) c^2 \; {\dot r_s(t)\over r_s(t)} 
=- {3\over2}\left(2 G_N m(t)\over r_s(t) c^2\right) {c^5\over G_N} \left(\dot r_s(t) \over c\right)
\nonumber\\
&=& {3\over2} P_* \left(2 G_N m(t)\over r_s(t) c^2\right)  \left(\dot r_s(t) \over c\right).
\end{eqnarray}
Now to prevent black hole formation we do want the compactness to be less than unity
\begin{equation}
{2 G_N m(t)\over r_s(t) c^2} < 1.
\end{equation}
Thence
\begin{equation}
P(t) < - {3\over2} P_*\left(\dot r_s(t) \over c\right).
\end{equation}
That is
\begin{equation}
\wp (t) < - {3\over2}\left(\dot r_s(t) \over c\right).
\end{equation}
But what if anything can we say about $\dot r_s(t)/c$? 
Naively one might wish to assert $|\dot r_s(t)|<c$, but we shall soon see that such a postulated constraint falls apart upon closer inspection.

%========================================================
\section{Nothing can travel faster than light?}\label{C:no-thing}
%========================================================

 It is a truism of special relativity that``nothing can travel faster than light'', or more precisely, as emphasized by both Taylor and Wheeler~\cite{Taylor-Wheeler}, and by Rothman~\cite{Rothman},  
 \break
 ``no \emph{thing} can travel faster than light''.  That is,  ``no \emph{physical object} can travel faster than light''. But does the location of the surface of an evaporating Schwarzschild constant density star
 (or by extension, the location of the surface of \emph{any} evaporating stellar model) qualify as a ``physical object''?
 This is a delicate question with model-dependent answers.
 
 %\clearpage
 Suppose one is dealing with a star made of baryons, and the only energy loss is due to photons: In such a situation the 4-velocity of the surface is determined by the average 4-velocity of the baryons in the immediate vicinity of the surface.  But the average of future-pointing timelike vectors is still a future-pointing timelike vector. So in this specific situation we certainly have $|\dot r_s(t)|<c$.
 Unfortunately there are very many other physical scenarios where such a simple argument does not apply. 
 
 \enlargethispage{30pt}
 Suppose now that one is dealing with a star whose outermost layers are being blown off explosively. 
 Some of the details of the explosion process are now important. The location of the surface $r_s(t)$ is now less clearly definable as a ``physical object'', it is simply a demarcation point between a more-or-less stable core and the now explosively dispersing former outer layers of the object in question. (A mathematical boundary is not necessarily a physical object.) A ``detonation wavefront'' has more in common with superluminal \emph{non-things}, such as relativistic scissors, a relativistic searchlight sweep, or relativistic oscilloscope writing speeds~\cite{Taylor-Wheeler, Rothman}; all of these phenomena share in common a certain delicate dependence on initial conditions. Whether or not the material in the vicinity $r=r_s(t)$ is about to explode depends on how close the the material in the vicinity of $r=r_s(t)$ is to some irreversible phase transition~\cite{collapse,Yunes} --- there is no \emph{a priori} need for a causal subluminal signal to propagate inwards to tell the star to explode. 
 
  %%% mathematical boundary =/= physics...

%========================================================
\section{Misner--Sharp quasi-local mass in general}\label{S:quasilocal}
%========================================================
 
 Let is now extend the discussion beyond the highly idealized evaporating version of Schwarzschild's constant density star.
 Given only spherical symmetry one can define a Misner--Sharp quasi-local mass $m(r,t)$.
 Given in addition a well-defined surface $r_s(t)$ this specializes to $m(t) = m(r_s(t),t)$. In terms of the average density $\bar \rho(t)$ one can without further loss of generality write~\cite{Misner-Sharp}
\begin{equation}
m(t) =  {4\pi\over 3} \; \bar \rho(t) \; r_s(t)^3. 
\end{equation}
So for the luminosity we now have
\begin{equation}
P(t) = -\dot m(t) =  -  m(t) \; { \dot{\bar \rho}(t)\over \bar \rho(t)}  -3 m(t) \; {\dot r_s(t)\over r_s(t)}.
\end{equation}
Now \emph{if} we assert $\dot{\bar \rho}(t) \geq 0$, \emph{then} we can deduce
\begin{equation}
P(t) \leq  -3 m(t) \; {\dot r_s(t)\over r_s(t)},
\end{equation}
and follow through by adapting the analysis presented above for the evolving Schwarzschild constant density star:
\begin{equation}
P(t) \leq  -{3\over2} \;{2G_N m(t) \over r_s(t) c^2} \;P_*\;  {\dot r_s(t)\over c } 
<  -{3\over2}  \; P_*  \; {\dot r_s(t)\over c }.
\end{equation}
Ultimately
\begin{equation}
\wp(t) \leq   -{3\over2}  {\dot r_s(t)\over c } .
\end{equation}
That is, \emph{if} we assume four conditions: (i) spherical symmetry, (ii) a nondecreasing average density, $\dot{\bar \rho}(t) \geq 0$, (iii) absence of horizons, $2G m/r_s <1$, \emph{and} (iv) 
subluminal motion of the surface, $|\dot r_s| <c$, \emph{then} the luminosity is indeed bounded $\wp < {3\over2}$. 

So to derive bounded luminosity requires some significant assumptions (beyond just invoking standard general relativity). The weakest of these assumptions, as argued above, is assuming $|\dot r_s| <c$ --- we really have no good physics reason for making this assumption. 
The nondecreasing condition on average density, $\dot{\bar \rho}(t) \geq 0$, is also somewhat questionable.  Certainly if the stellar object ever completely disperses one should expect
${\bar \rho}(t) \to 0$,  requiring $\dot{\bar \rho}(t) < 0$ for at least part of the object's history.

%==================================
\section{Discussion and conclusions}
%==================================

In this essay we have reviewed and re-analyzed the maximum luminosity conjecture, with a view to clarifying just how generic such a conjecture might actually be. We have seen that within the framework of general relativity the \emph{exterior} spacetime places no physical constraint on the total luminosity --- the only conceivable way in which one might place a bound on the total luminosity is by investigating the \emph{interior} spacetime geometry of the source; and that is a rather model-dependent project with results that seem to be less than universal.

%==================================
\section*{Acknowledgments}
%==================================
AJ was indirectly supported by the Marsden Fund, via a grant administered by the Royal Society of New Zealand.\\
MV was directly supported by the Marsden Fund, via a grant administered by the Royal Society of New Zealand.

%========================================================
%========================================================

%========================================================
\end{document}